\documentclass[traditabstract,twocolumn]{aa}
 \usepackage{txfonts}
 \usepackage{pslatex}
 \usepackage{caption}
 \usepackage{graphicx}
 \usepackage{dcolumn}
 \usepackage{bm}
 \usepackage{subfig}
 
 \newcommand\simlt{\lower.5ex\hbox{$\; \buildrel < \over \sim \;$}}
 \usepackage{natbib}
 \bibpunct{(}{)}{;}{a}{}{,} 

\newcommand{\cm}{\,{\rm cm}}

\newcommand{\uG}{\,\mu{\rm G}}

\newcommand{\gray}{$\gamma$-ray}
\newcommand{\grays}{$\gamma$-rays}

\newcommand{\Fermi}{{\sl Fermi}}

\begin{document}

\title{$\Fermi$ Large Area Telescope observations of the supernova remnant HESS J1731-347}
\author{Rui-zhi Yang\inst{1,4}
 \and Xiao Zhang\inst{2}
 \and Qiang Yuan\inst{3}
  \and Siming Liu\inst{1}
}
\institute{Key Laboratory of Dark Matter and Space Astronomy, Purple Mountain Observatory, Chinese Academy of Sciences, Nanjing, 210008, China;  ryang@mpi-hd.mpg.de, liusm@pmo.ac.cn
\and Department of Astronomy, Nanjing University, Nanjing 210093, China
\and Key Laboratory of Particle Astrophysics, Institute of High Energy Physics, Chinese Academy of Sciences, Beijing 100049, China
\and 
}

\date{Received 2013 / Accepted 2014}
\abstract
{
{\it Context.}
HESS J1731-347 has been identified as one of the few TeV-bright shell-type supernova remnants (SNRs). These remnants are dominated by nonthermal emission, and the nature of TeV emission has been continuously debated for nearly a decade.
\\
{\it Aims.} We carry out the detailed modeling of the radio to $\gamma$-ray spectrum of HESS J1731-347 to constrain the magnetic field and energetic particles sources, which we compare with those of the other TeV-bright shell-type SNRs explored before.\\
{\it Methods.}
Four years of data from {\it Fermi} Large Area Telescope (LAT) observations for regions around this remnant are analyzed, leading to no detection correlated with the source discovered in the TeV band. The Markov Chain Monte Carlo method is used to constrain parameters of one-zone models for the overall emission spectrum.
\\
{\it Results.}
Based on the 99.9\% upper limits of fluxes in the GeV range, one-zone hadronic models with an energetic proton spectral slope greater than 1.8 can be ruled out, which favors a leptonic origin for the $\gamma$-ray emission, making this remnant a sibling of the brightest TeV SNR RX J1713.7-3946, the Vela Junior SNR RX J0852.0-4622, and RCW 86.
The best-fit leptonic model has an electron spectral slope of 1.8 and a magnetic field of $\sim$30 $\mu$G, which is at least a factor of 2 higher than those of RX J1713.7-3946 and RX J0852.0-4622, posing a challenge to the distance estimate and/or the energy equipartition between energetic electrons and the magnetic field of this source. A measurement of the shock speed will address this challenge and has implications on the magnetic field evolution and electron acceleration driven by shocks of SNRs.
}
\titlerunning{}
\maketitle

\section{Introduction}
Shocks of supernova remnants (SNRs) have been suggested as the major acceleration site of Galactic cosmic rays for several decades. The discovery of TeV $\gamma$-ray emission from the shell-type SNR RX J1713.7-3946 strengthens this proposal \citep{A2004}, however, the nature of the TeV emission has been a matter of debate ever since \citep[e.g.,][]{kw2008,Yuan2011}. More extensive observations uncover a class of shell-type SNRs with the overall emission dominated by energetic particles. Two prototypes are SNR RX J1713.7-3946 and the Vela Junior SNR RX J0852.0-4622  \citep{hess2007,A2009}. Recent {\it Fermi}-LAT observations of these two sources have revealed very hard spectra in the GeV range \citep{T2011, A2011}, favoring a leptonic scenario for the $\gamma$-ray emission \citep{Y2012}. Interestingly, a detailed magnetohydrodynamics (MHD) modeling of SNR RX J1713.7-3946 shows that the energy density of electrons above 1 GeV is equal to that of the magnetic field \citep{YL2013}, which lends further strength to the leptonic models.  Although RCW 86 has relatively more complex structures in different wave bands and evident thermal emission \citep{2006ApJ...648L..33V, 2009Sci...325..719H}, \citet{2012A&A...545A..28L} recently showed that the characteristics of its nonthermal emission are very similar to SNR RX J1713.7-3946 and RX J0852.0-4622.

HESS J1731-347 was first seen as an unidentified source with HESS observations \citep{hess2008}. Later observations confirmed its similarity to SNR RX J1713.7-3946 \citep{Tian2010, hess2011, Bam2012}. A preliminary survey of {\it Fermi}-LAT observations revealed a nearby source, 1FGL J1729.1-3452c, whose GeV flux is treated in a previous study  as an upper limit for emission from HESS J1731-347  \citep{hess2011}. In this paper, we carry out a detailed analysis of {\it Fermi}-LAT observations of this source. Given the relatively complete spectral coverage of these sources, \citet{Yuan2011} first used the Markov Chain Monte Carlo (MCMC) method to constrain parameters of different emission models for SNR RX J1713.7-3946 \citep{lewis2002}. We carry out a similar study here to explore the nature of the $\gamma$-ray emission.

Although the multiwavelength spectrum of these sources can be obtained directly from observations , it is very difficult to determine their distance and age \citep{A2007A&A, 2008K, hess2011}, which makes further exploration of their nature challenging. The distance of an SNR is usually constrained by observations of emission or absorption lines from molecular clouds along the line of sight whose distance can be inferred from the Doppler shift of the line and the spiral structure of the Milky Way galaxy \citep[e.g.,][]{2005E, Tian2008, Tian2010}. Further evidence of interaction of SNRs with molecular clouds can then be used to determine the distance of SNRs. A distance estimate can also be obtained from the X-ray absorption column depth inferred with spectral measurements made in the X-ray band. \citep{hess2011}. The age of SNRs can be inferred from the distance and angular size based on some models of the evolution of the shock in the interstellar medium (ISM) and can be cross-checked with other related observations \citep{1997W}.

Given the similarity of these sources, especially the dominance of emission by energetic particles produced by shocks, the emission spectrum is expected to have some universal time evolution history and therefore might be correlated with the distance and/or the angular size \citep{N2012, 2013D}. {  Based on the current distance estimate of 3.2 kpc, 1 kpc, 1 kpc, and 2.5 kpc for HESS J1731-347, J0852.0-4622, J1713.7-3946, and RCW 86 ,respectively, these four remnants have comparable luminosity in the TeV range \citep{hess2011}. For the same distance estimates, however, the X-ray luminosity of HESS J1731-347 is more than 10 times higher than that of the Vela Junior SNR RX J0852.0-4622 and about 2 times higher than those of SNR RX J1713.7-3946 and RCW 86. The diameter of the shell of HESS J1731-347, J0852.0-4622, J1713.7-3946, and RCW 86 are about 27 pc, 34 pc, 17.4 pc, and 30 pc, respectively. These differences need to be explained in the context of shock evolution and the related magnetic field amplification and particle acceleration processes \citep{2012G, R2013, YL2013}.}

Our data analysis is presented in Section 2, followed by an MCMC exploration of the parameter space of different emission models for the overall spectrum (Section 3). In Section 4, we conduct a comparative study of these four remnants and reveal a challenge to the leptonic model for HESS J1731-347. The conclusion is drawn in Section 5.

\section{Data analysis}
\label{data}

We selected nearly four years of data (MET 239557417 - MET 355051421) from the {\it Fermi}-LAT observations for regions around the shell of the SNR HESS J1731-347 and used the standard LAT analysis software (v9r27p1)
\footnote{http://fermi.gsfc.nasa.gov/ssc}. To study the $\gamma$-ray morphology in this region, only events with energy above 2 GeV were used so that the point spread function (PSF) is sharp enough to disentangle multiple spatial components. The region-of-interest (ROI) was selected to be a $10^ \circ \times 10^ \circ$ square centered on the position of HESS J1731-347.  To reduce the effect of the Earth albedo background, time intervals when the Earth was appreciably in the field-of-view (FoV) \footnote{That is when the center of the FoV is more than $52^ \circ$ from zenith as well as time intervals when parts of the ROI are observed at zenith angles $> 100^ \circ$.} were excluded from the analysis. The spectral analysis was performed based on the P7v6 version of post-launch instrument response functions (IRFs).  Both the front and back converted photons were selected.

The Galactic and isotropic diffuse models provided by the {\it Fermi} collaboration \footnote{Files: gal\_2yearp7v6\_v0.fit and iso\_p7v6source.txt available at \\ http://fermi.gsfc.nasa.gov/ssc/data/access/lat/BackgroundModels.html} were used in the analysis. We included 2FGL sources, and allowed parameters for point sources within $3^{\circ}$ of HESS J1731-347 to vary. The residual map of the inner $5^{\circ}$ is shown in Figure 1, where the HESS contours are also over-plotted. From the residual map we see that there is no evident excess in the area of the shell of the SNR, however, a strong excess can be seen toward the southwest of the shell.  This excess is near the 1FGL {\it Fermi} point source 1FGL source J1729.1-3452c, but is absent in the 2FGL catalog. To locate this source we ran {\it gtfindsrc} and found that the best-fit position is at ($R.A.(J2000) = 262.22^{\circ}, Dec.(J2000) = -35.00^{\circ}$), which is about $0.7^{\circ}$ away from the center of the SNR. Considering the fact that the angular resolution of {\it Fermi} LAT above 2 GeV is about $0.2^{\circ}$, we argue that this source may not be associated with the SNR itself and treat it as a background point source.

This point source has a flux of $1.6\times 10^{-9}~\rm ph\ cm^{-1} s^{-1}$ and a TS value of about 20 above 2 GeV. The {\it null hypothesis} of our model includes this point source and all the other sources in the 2FGL catalog. The significance of the SNR can be derived by comparing the likelihood  in both the {\it null hypothesis} ($\mathcal{L}_0$) and  the tested model including the SNR ($\mathcal{L}_1$). The TS value is defined as
\begin{equation}
TS=-2\log(\mathcal{L}_0/\mathcal{L}_1)\,.
\end{equation}
We used both a point source and a spatial template generated with the H.E.S.S image as the spatial model of the SNR in the tested models and found that both of them cannot give a positive TS value. This is compatible with the fact that there is no excess in this region in the residual map.

\begin{figure*}[htb]
\centering
\includegraphics[width=80mm,angle=0]{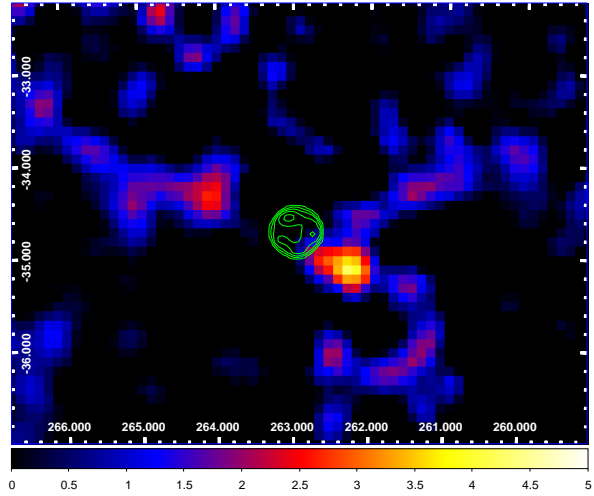}
\caption{Residual counts map above 2 GeV, derived by subtracting the best-fit model map from the counts map. The unit of the map is counts per pixel while the pixel size is $0.1^{\circ}\times0.1^{\circ}$. Green contours show the HESS image of the SNR at 10, 30, 50, 70, 90\% of the corresponding peak value of 134 counts per $0.04^{\circ}\times0.04^{\circ}$. The other nearby TeV source HESS J1729-345 is not shown here. The new point source with a TS value 20 mentioned in the text is not subtracted in this residual map.} 
\end{figure*}

To get the upper limits of the fluxes of the SNR we added an extended source  consistent with the H.E.S.S observations, i.e., a uniform disk with a radius of $0.3^{\circ}$  centered at ($R.A.(J2000) = 262.97^{\circ}
, Dec.(J2000) = -34.75^{\circ}$ ), as well as the point source mentioned above. Three independent energy bins were adopted to calculate the upper limits.   Although we used the data above $2~ \rm GeV$ to investigate the morphology due to the better angular resolution in higher energies, all the data above $100~\rm MeV$ were used to derive the flux upper limits with the python tool $\it UpperLimits$,  which assumes that a decrease of the log-likelihood by 4.8 gives the 99.9\% upper limits.   We modeled the emission from HESS J1731-347 using a power-law spectral distribution with two free parameters: the flux and the spectral index. When the spectral index is fixed to $2$,
the 99.9\% upper limits in these energy bins are shown in Figure 2 corresponding to $4.0 \times 10^{-8}$, $1.2 \times 10^{-9}$, and $1.7 \times 10^{-10}~\rm ph\ cm^{-1} s^{-1}$ at 0.1--1 GeV, 1--10 GeV, and 10--100GeV, respectively. We also tried the models with the spectral index fixed at 1.5, and 3, and the resultant differences are less than 10\%.

\begin{figure*}[htb]
\centering
\includegraphics[height=54mm,angle=0]{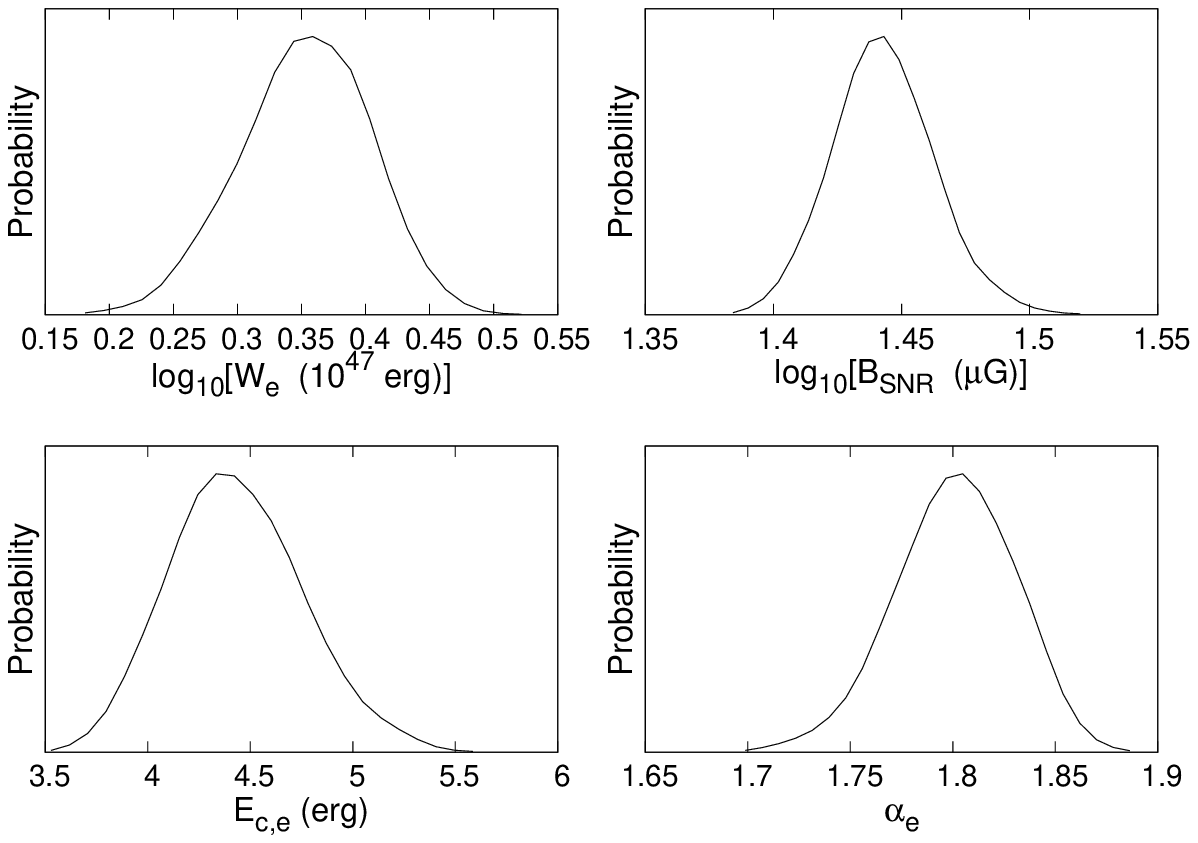}
\includegraphics[height=55mm,angle=0]{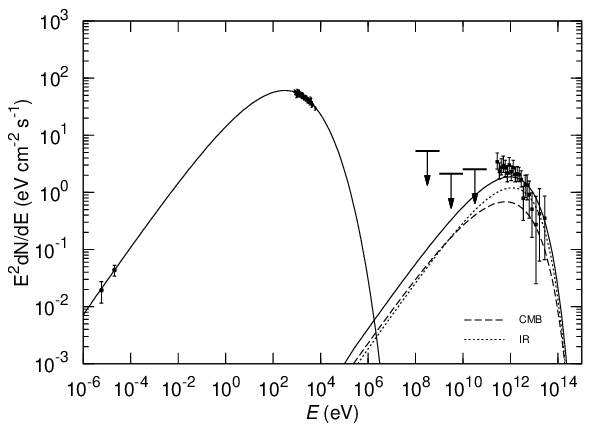}
\caption{Leptonic scenario for $\delta_{e}$ = 0.5. Left: 1D probability distribution of the model parameters. Right: the best-fit to the SED derived from observations in the radio \citep{Tian2008}, X-rays \citep{hess2011}, which have been scaled upward by a factor of 2.75 due to the incomplete spatial coverage of the XMM images, and $\gamma$-rays \citep{hess2011}. The corresponding model parameters are listed in the Table~\ref{tab:para}.}
\label{fig:lep}
\end{figure*}

\section{The MCMC modeling of the emission spectrum}
\label{dis}

To study the origin of the \grays , we use a simple one-zone static model to fit the multi wavelength spectral data assuming a distance of 3.2 kpc \citep{hess2011} and use the MCMC technique \citep[see, e.g., ][and references therein]{Yuan2011} to constrain the model parameters.
In this model, the spatially integrated energy spectrum of accelerated particles (electrons and protons) in the emission region is assumed as
\begin{equation}
dN/dE \propto E^{-\alpha} {\rm exp}[-(E/E_{c})^{\delta}]\,,
\end{equation}
 where $E, \alpha, and E_{c}$ are the particle energy, power-law spectral index, and the cut off energy, respectively, and $\delta$ describes the sharpness of this cutoff. The normalization is determined from the total kinetic energy of particles above 1 GeV, $W$ \footnote{The subscript ``$e$" and ``$p$" will be employed to differentiate the parameters between electrons and protons.}. The emission spectra from such energetic particle distributions can be readily obtained with the relevant emission mechanisms for a uniform emission region with a constant magnetic field \citep{Sturner1997, Kelner2006}.  This simple emission model  neither considers the electron energy losses, which will affect the energy content of energetic electrons, nor addresses the detailed process of particle acceleration, which may be constrained, however, by studying the characteristics of the model parameters \citep{2010A&A...517L...4F}. And currently observations do not justify more complex models \citep{2012A&A...545A..28L}.

We first consider the pure leptonic scenario in which the \grays\ and the radio to X-ray emission are generated by relativistic electrons via the inverse Compton (IC) scattering on the background radiation field and the synchrotron process \citep{Sturner1997}. For this source, besides the cosmic microwave background (CMB), an infrared (IR) seed photons with an energy density of 1 eV$\cdot\cm^{-3}$ and a temperature of 40 K  is also included \citep{hess2011}. In this picture, there are five parameters, including $W_{e},\ E_{c,e},\ \alpha_{e}, \ \delta_{e}$ and $B_{\rm SNR}$, which is the average magnetic field in the SNR. If all of them are set free in the MCMC routine, however, $E_{c,e},\ \alpha_{e}$,  and $\delta_{e}$ do not converge very well with multiple peaks because of the low quality of the X-ray and/or \gray\ data that constrain parameters of the spectral shape \footnote{Note that the GeV upper limits are not used in the MCMC calculation.}. Therefore, instead of setting $\delta_{e}$ free in the MCMC fit, we take some typical values of 0.5 \citep{Liu2008}, 0.6 \citep{Li2011} and 1.0 \citep{2012A&A...545A..28L} for this parameter and explore constraints on the other parameters.
The one-dimensional (1D) probability distribution and the best-fit spectral energy distribution (SED) for the $\delta_{e}=0.5$ are displayed in Figure~\ref{fig:lep} \footnote{For the other two cases in the leptonic scenario, the spectra are similar to this one.}. In the 1D probability distribution, the peaks corresponds to the best-fit model parameters which together with 1$\sigma$ statistical uncertainties are listed in Table~\ref{tab:para}. The $\chi^{2}$ values of the best-fit model can be found in Table~\ref{tab:chi2}.

\begin{table*}[htb]
\caption{Fitting parameters}
\label{tab:para}
\begin{tabular}{lcc|cccccc}
\hline\hline
Model& $\delta_{e}$& $\alpha_{p}$& $\alpha_{e}$&$B_{\rm SNR}$& $W_{e}$ & $E_{c,e}$ & $W_{p}$ & $E_{c,p}$ \\
         &    &      &              & ($\uG$)      &($10^{47}$ erg) & (TeV)   &($10^{50}$ erg)& (TeV)    \\ \hline
Leptonic & 0.5& ---  & $1.80^{+0.03}_{-0.03}$ & $27.74^{+1.23}_{-1.18}$ & $2.25^{+0.27}_{-0.24}$ & $2.77^{+0.20}_{-0.20}$ &  ---& --- \\
& 0.5& ---  & $\left(1.80^{+0.05}_{-0.08}\right)^*$ & $\left(16.63^{+1.78}_{-1.23}\right)$ & $\left(5.02^{+1.21}_{-1.31}\right)$ & $\left(3.54^{+0.58}_{-0.54}\right)$ &  ---& --- \\
         & 0.6& ---  & $1.88^{+0.03}_{-0.03}$ & $27.33^{+1.29}_{-1.26}$ & $2.09^{+0.26}_{-0.23}$ & $6.26^{+0.40}_{-0.41}$ &  ---& ---  \\
         & 1.0& ---  & $2.02^{+0.02}_{-0.02}$ & $26.32^{+1.32}_{-1.25}$ & $1.77^{+0.24}_{-0.21}$ & $23.62^{+1.32}_{-1.25}$ &  ---& ---  \\ 
Hadronic & 1.0& 2.0 & $2.02^{+0.02}_{-0.02}$ & $73.54^{+97.76}_{-39.46}$ & $0.34^{+0.76}_{-0.25}$ & $14.79^{+6.19}_{-6.21}$ & $3.77^{+0.58}_{-0.49}$ & $43.47^{+15.92}_{-11.61}$ \\
         & 1.0& 1.8 & $2.02^{+0.02}_{-0.02}$ & $72.53^{+178.98}_{-57.61}$ & $0.36^{+1.86}_{-0.31}$ & $14.25^{+9.68}_{-12.65}$ & $2.24^{+0.51}_{-0.44}$ & $28.55^{+14.81}_{-9.43}$ \\
\hline
\end{tabular}\\

{Note.}\\
{\rm *}: Numbers in parentheses are for the model with the CMB as the only background radiation. 
\end{table*}

In general, the three leptonic cases with different $\delta_{e}$ values have very similar results. As $\delta_{e}$ increases, all parameters except for $E_{c,e}$, which has one order of magnitude variation from $\sim 2.5$ to $\sim 25$ TeV, change very little. The dramatic increase of $E_{c,e}$ is understandable because a sharper cutoff (a higher value of $\delta_e$) requires a higher cutoff energy to have sufficient energetic electrons to produce the observed X-ray and \gray\ fluxes. The spectral slope also increases slightly with the increase of $\delta_e$. For the soft background photon field adopted here, the average magnetic field in the SNR is confined in the range of 25 -- 30 $\uG$ , which is about 2 times higher than those found in SNR RX J1713.7-3946 and the Vela Junior SNR RX J0852.0-4622 \citep{Liu2008} due to the higher X-ray to $\gamma$-ray flux ratio.  The fitting parameters for the model with the CMB as the only background photon field are also given in Tables \ref{tab:para} and \ref{tab:chi2}. In this case, the total energy of electrons will be larger by a factor of $\sim2$, and the magnetic field is found to be $\sim17$ $\mu$G, which is $\sim1.5$ times smaller and can be considered as a lower limit to the $B$ field. This magnetic field is similar to the 15-25 $\mu$G $B$ field that \citet{2012A&A...545A..28L} found in RCW 86.

The overall $\chi^{2}$ and that of the \gray\ data increase significantly with the increase of $\delta_e$, however, this does not mean that the TeV data is better-fitted with a shallower high-energy cutoff (corresponding to a lower value of $\delta_e$).
Fitting the TeV data alone with $W_{e},\ E_{c,e}$, and $\delta_{e}$ as free parameters and $\alpha_{e}$ at $1.8$ and 2.0, we got similar results with a value of $\sim$6.2 for the $\chi^{2}$ of the \gray\ data and $\delta_{e}=0.89\pm0.19$.
In addition, for all cases in the leptonic scenario, the \gray\ data with energy less than $\sim$1 TeV still can not be fitted very well (as an example, see the right panel of Figure~\ref{fig:lep}). This is the major limitation of the simple leptonic model.
Next, we discuss the scenario where the hadronic process is responsible for the origin of the \gray\ emission.

\begin{table*}[htb]
\caption{Best-fit $\chi^{2}$ value for each set of parameters.}
\label{tab:chi2}
\begin{tabular}{lcc|cccc}
\hline\hline
Model&$\delta_{e}$&$\alpha_{p}$& Radio &  X-ray  &   TeV   &  Reduced   \\ \hline
Leptonic  &   0.5   &  ---   &   0.02  &  26.11  &  19.28  &  45.42/61  \\
          &   0.5   &  ---   &   (0.06)$^*$  &  (25.94)  &  (21.07)  &  (47.07/61)  \\
          &   0.6   &  ---   &   0.06  &  25.78  &  22.24  &  48.10/61  \\
          &   1.0   &  ---   &   0.19  &  25.97  &  31.88  &  58.04/61  \\ 
Hadronic  &   1.0   &  2.0  &   0.23  &  39.24  &   5.69  &  45.16/59  \\
          &   1.0   &  1.8  &   0.32 &       28.88&       5.53&       34.74/59  \\
\hline
\end{tabular}\\

{  Note.}\\
{\rm *}: Numbers in parentheses are for the model with the CMB as the only background radiation.
\end{table*}

\begin{figure*}[htb]
\centering
\includegraphics[height=60mm,angle=0]{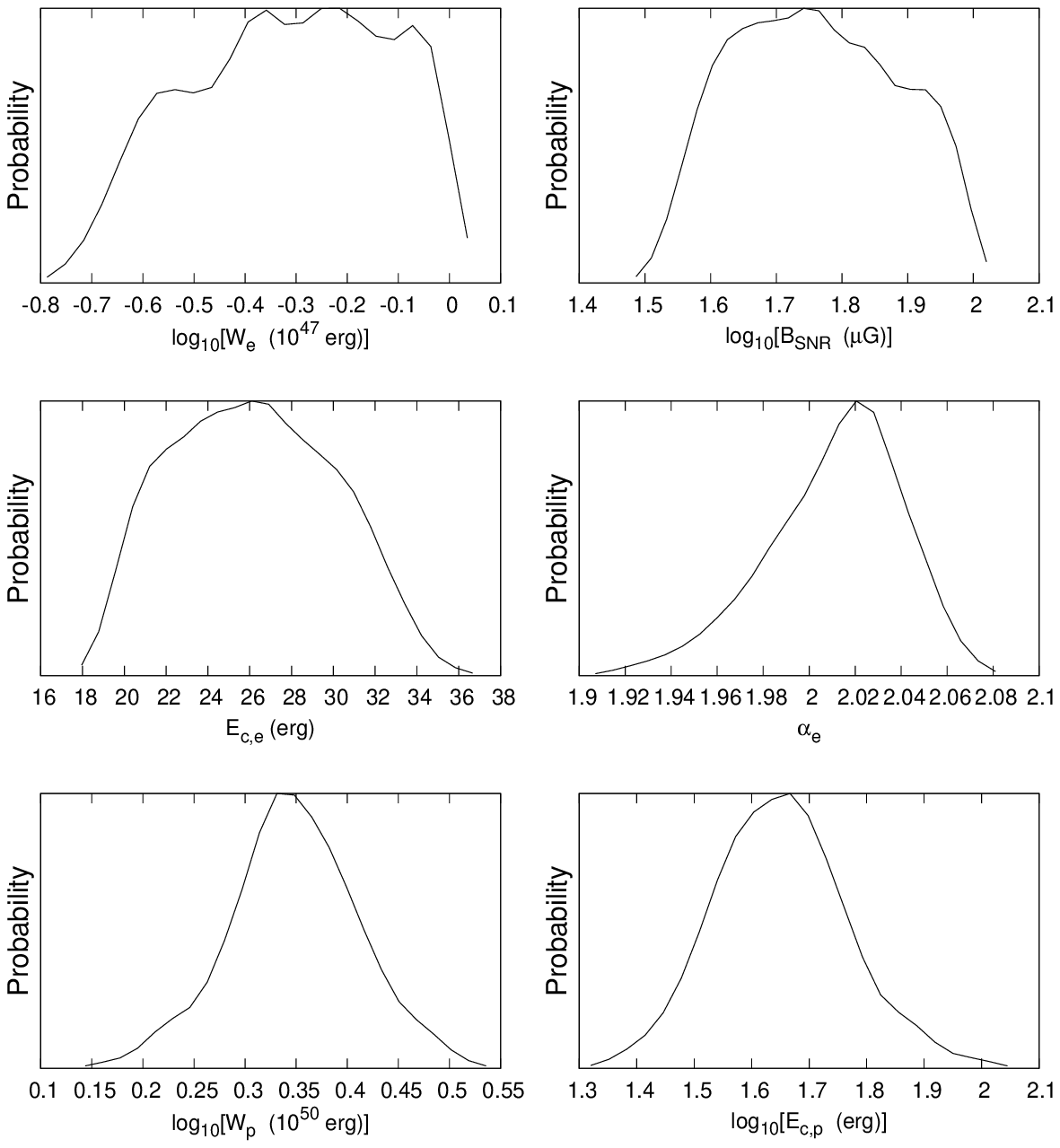}
\includegraphics[height=65mm,angle=0]{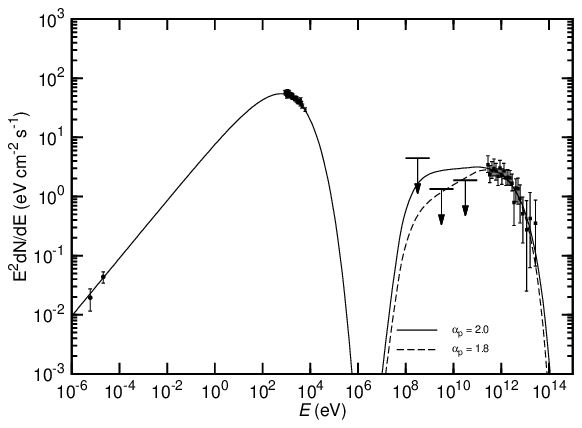}
\caption{Same as the right panel in Figure~\ref{fig:lep} but for the hadronic scenario. In the \gray\ band, the solid and dashed lines indicate the cases with the spectral index of protons $\alpha_{p}=2.0\ {\rm and}\ 1.8$. The density of the ambient medium is assumed to be 1 cm$^{-3}$.}\label{fig:had}
\end{figure*}

In the hadronic picture, the \grays\ come from the decay of neutral pions that are produced through the inelastic collisions of energetic hadrons (mainly protons) with nuclei in the background plasma \citep[we adopted the parameterized method given in][]{Kelner2006}. In this scenario, besides parameters describing the energy distribution of protons, one also needs to know the density of the ambient medium, $n_{0}$. Because of lack of observational constraint on the density, it was fixed at 1 $\cm^{-3}$ here following \citet{hess2011}.  It should be noted that because of the low density  of the ambient medium, we neglected the nonthermal electron bremsstrahlung in the modeling. 

Because the spectral parameters of protons are only constrained by the TeV data in the MCMC fit, there is a strong degeneracy with large errors for the best-fit model parameters \citep{Yuan2011}. To reduce the number of free parameters, we fixed $\delta_e=\delta_p$ at 1.
To take into account the 3$\sigma$ upper limits of the GeV data, which were not included in the MCMC fit,
$\alpha_{p}$ was chosen as a prior in the MCMC fit. 
In combination with the parameters in the leptonic models, in total there are six free parameters: $W_{e}$, $B_{\rm SNR}$, $E_{c,e}$, $\alpha_{e}, W_{p}$, and $E_{c,p}$ in the MCMC fit of the hadronic model. The best-fit SED is shown in Figure~\ref{fig:had}. The corresponding parameters and $\chi^{2}$ value are listed in Table~\ref{tab:para} and Table~\ref{tab:chi2}, respectively.
As expected, the $\chi^2$ value of the \gray\ data is improved significantly in the hadronic model ( Table~\ref{tab:chi2}).
In order to be consistent with the Fermi upper limit, however, a proton spectral slope of $\alpha_{p} \lesssim 1.8$ is needed.
This value of the proton spectral slope is definitely flatter than the slope, 2.1 -- 2.4, derived from observations of the local cosmic ray spectrum \citep{Gabici2009} and is also slightly harder than 2.0 predicted by the test particle version of the diffusive shock acceleration model (DSA). Of course, if one considers the nonlinear effects, the spectral index can be reduced to 1.5, which is no longer in conflict with the results. 
In this scenario, 
the free hadronic parameters are completely independent of the electronic parameters and are directly constrained by the \gray\ data.The hadronic parameters have converged very well despite relatively large 1$\sigma$ errors.
The electronic parameters, on the other hand, do not converge very well due to lack of constraint from the \gray\ data.
All electronic parameters except $\alpha_{e}$ cannot be well constrained with large 1$\sigma$ errors. 


\section{Discussion}

The above study puts HESS J1731-347 in the same category of TeV-bright shell type SNRs RX J1713.7-3946, J0852.0-4622, and RCW 86 with the TeV emission likely originated from the inverse Comptonization of background photons by energetic electrons \citep{Y2012}. A detailed examination of the model parameters, however, reveals a challenge to the leptonic model for HESS J1731-347 and RCW 86, especially in light of the recently noticed energy equipartition between energetic electrons and the magnetic field in RX J1713.7-3946 \citep{YL2013}. Table 3 lists key parameters for these four sources.

\begin{table*}[htb]
\caption{Comparison of HESS J1713.7-3946, J0852.0-4622, RCW 86, and J1731-347}
\label{tab:snrs}
\begin{tabular}{l|cccccc}
\hline\hline
Name             &$F_\gamma^{\rm a}$
/$F_X^{\rm b}$
& Diameter$^{\rm c}$
& $B^{\rm d}$
& $W_e^{\rm d}$ 
& $W_B$\\

            &$10^{-10}{\rm erg\ s^{-1} cm}^{-2}$
              & pc 
            &  $\mu$G & $10^{47}$ erg & $10^{47}$ erg \\ \hline
RX J1713.7-3946            &   0.68/5.4
&   17.4($D/1.0$) 
&  12 & 3.9($D/1.0$)$^2$
& 4.0($D/1.0$)$^3$($f/0.87$)$^{\rm e}$ \\
RX J0852.0-4622              &   0.66$^{\rm e, f}$/0.83
&   34($D/1.0$)
&  9.4 & 10($D/1.0$)$^2$  
& 10($D/1.0$)$^3$($f/0.49$)$^{\rm f}$\\
HESS J1731-347    &   0.09/1.0$^{\rm a}$
&     27(D/$3.2$)
& 28 & 2.3($D/3.2$)$^2$ 
& 85($D/3.2$)$^3$($f/0.9$)$^{\rm a}$\\
 RCW 86    &   0.09/0.41$^{\rm g}$
&     30(D/$2.5$)
& 25 & 9.3($D/2.5$)$^{2{\rm *}}$ 
& 91($D/2.5$)$^3$($f/0.88$)$^{\rm h}$
\\
\hline
\end{tabular}\\

{  Note.}\\
$F_\gamma$: $\gamma$-ray flux in the 1-30 TeV range; \\
$F_X$: X-ray flux in the 2-10 keV range; \\
$D$: Distance in kpc; \\
$W_B$: the total energy of magnetic field in the emission region; \\
$f$: Volume filling factor of the emission region.\\
{\rm *}: The low-energy cutoff has been set at 1 GeV. \\
{  References.} \\
a: \citet{hess2011}; b: \citet{N2012}; c: \citet{R2013}; d: \citet{Liu2008}; e: \citet{A2007A&A}; f: \citet{hess2007}; g: \citet{2012A&A...545A..28L}; h: \citet{2009ApJ...692.1500A}.
\end{table*}

Although RX J0852.0-4622 is 2 times larger in angular size and 6 times lower in X-ray flux than RX J1713.7-3946, their TeV fluxes and radial thickness of the emission regions are comparable \citep{hess2007, A2007A&A}. Assuming a distance of 1 kpc for both sources, these differences may be naturally explained with an older age of RX J0852.0-4622 expanding in a lower density environment, leading to a relatively larger size, and a lower magnetic field and X-ray luminosity \citep{N2012}. Adopting the same background photon field as that at 7.5 kpc from the Galactic center in the Galactic plane \citep{2006ApJ...648L..29P}, a detailed spectral fit by \citet{Liu2008} gives the total energy of electrons above 1 GeV of $10^{48}$ erg and $3.9\times 10^{47}$ erg, and a mean magnetic field of 12 $\mu$G and 9.4 $\mu$G for RX J0852.0-4622 and RX J1713.7-3946. The TeV observations \citep{hess2007, A2007A&A} show that the radial brightness profile is consistent with a uniform emitting shell with a thickness about 20\% and 50\% of the radius for RX J0852.0-4622 and RX J1713.7-3946, giving rise to a volume filling factor of the emission region of 0.49 and 0.87.
The total energy of the magnetic field is equal to that of the energetic electrons within a few percent for both sources, which is consistent with the recent detailed MHD modeling of SNR RX J1713.7-3946 by \citet{YL2013}. Even considering the uncertainties in the distance estimate of these two sources, the above picture does not change dramatically except that the energy partition between the $B$ field and energetic electrons will be different. For example, for a shorter distance of 750 pc to RX J0852.0-4622 derived by \citet{2008K}, the magnetic field energy will be comparable to that of RX J1713.7-3946 at 1 kpc and the total energy of energetic electrons is $\sim 40$\% higher than that of RX J1713.7-3946. For a distance of 200 pc to RX J0852.0-4622 as studied by \citet{hess2007}, however, it will be smaller (hence younger) than RX J1713.7-3946 with much less energy in the magnetic field and energetic electrons.

The TeV luminosity of HESS J1731-347 is comparable to the two other remnants, assuming a distance of 3.2 kpc, however, the X-ray luminosity of HESS J1731-347 is two times higher and its s diameter $\sim50$\% larger than RX J1713.7-3946. The spectrum and size of HESS J1731-347 clearly do not follow the trend outlined by the two SNRs discussed above. Adopting a similar background photon field, the leptonic model gives a mean magnetic field of $\sim28$ $\mu$G. For an estimated volume filling factor of the emission region of 0.9 \citep{hess2011}, the total energy of the magnetic field is more than 35 times greater than that of energetic electrons, which is distinct from the other two sources. To bring the magnetic field and energetic electrons closer to an energy equipartition, one may consider the leptonic model with the CMB as the only background photon field. The best-fit model has $W_e = 5.0\times10^{47}$ erg and $B = 17\ \mu$G, corresponding to a total magnetic field energy of $3.1\times 10^{48}$ erg, which is still more than 6 times greater than $W_e$.

The synchrotron energy loss timescale of $\sim 10$ TeV electrons 
in a magnetic field of $\sim30$ $\mu$G 
is about 1400 years, which should be shorter than the age of the remnant. Otherwise, the mean shock speed will be about 9000 km/s, which is too high for a remnant with a radius of 13.5 pc. The synchrotron energy loss process then plays a role in the formation of high-energy cutoff of the electron distribution, which is different from the other two remnants, where the synchrotron energy loss timescale of the highest energy electrons is longer than the age of the remnants.  For a distance of 3.2 kpc, \citet{hess2011} estimated an age of 2500 years, which is significantly older than the synchrotron energy loss timescale of $\sim 10$ TeV electrons in a magnetic field of $\sim30$ $\mu$G. This may cause the low total energy of energetic electrons in this source.
The acceleration timescale of relativistic electrons via the Fermi process is greater than $r_g c/U^2$, where $r_g$ is the electron gyro-radius and $U$ is the speed of the scattering agent. Assuming the acceleration timescale of $\sim 10$ TeV electrons is comparable to or shorter than its synchrotron energy loss timescale, then we have $U> 260\ (E/10{\rm TeV})(B/30\mu{\rm G})^{1/2}{\rm km/s}$. A shock speed of a few thousand km/s is capable of accelerating electrons to $\sim 10$ TeV.

Alternatively, the energy equipartition between the magnetic field and energetic electrons may only hold for strong shocks in young SNRs. As the shock slows down, electron injection may stop while the magnetic field can be continuously amplified by the turbulent motion in the shock downstream \citep{2012G}, however, this model does not explain the fact that the Vela Junior remnant has the largest size with the weakest magnetic field. It appears that environmental factors must play a role in the difference of these sources. HESS J1731-347 is closer to the Galactic center than the other two. It is possible that its surrounding ISM has a higher magnetic field and density than the other two sources so that the magnetic field is amplified to a higher value and the shock slows down faster \citep{2012G}. But a shock speed of a few thousand km/s is still needed since continuous acceleration of high-energy electrons is necessary in this leptonic model given the short synchrotron energy loss timescale of $\sim 10$ TeV electrons.

The deviation from energy equipartition between magnetic field and energetic electrons in HESS J1731-347 is mainly determined by its relatively higher X-ray to $\gamma$-ray flux ratio in the leptonic scenario for the TeV emission. Different choices for the background photon field do not change the fact that the magnetic field has a much higher total energy than energetic electrons.  Since the magnetic field energy scales as the cube of the distance while the total energy of energetic electrons scales as the square of the distance, if the distance to HESS J1731-347 is within 500 pc, the magnetic field and energetic electrons can be in energy equipartition, and its strong magnetic field may be attributed to its relatively young age and small size \citep{N2012}. This implies a high shock speed with an angular expansion rate of $\sim 1\arcsec$ per year, which may be tested with high resolution X-ray observations \citep{2008K}. The age of the remnant will be shorter than 1400 years in this case so the synchrotron energy loss does not affect the distribution of energetic electrons. This scenario, however, has difficulties in accounting for the high absorption column depth inferred from the X-ray spectral measurement \citep{hess2011}.

The discussion above invokes two competing models for the evolution of magnetic fields in SNRs. Based on MHD simulations, \citet{2012G} show that the magnetic field stress in the shocked ISM increases continuously with the evolution of the remnants. Higher shock speed in the earlier stage of the remnant evolution makes the field stress grow faster. The field stress itself also depends on its value in the preshocked ISM. Their simulations assumed a uniform turbulent field in the ISM. Modification of the radial dependence of the field stress in the ISM is not expected to change the conclusion that the field stress increases continuously with radius since the field is amplified by the turbulent motion and the field dissipation is weak. Their simulations did not consider the field evolution in the ejector. The field stress in the ejector is expected to decrease rapidly due to the expansion.
In the shock model proposed by \citet{N2012}, the field energy density is assumed to be proportional to the thermal energy density in the shock downstream. The field stress then decreases monotonically with the expansion of the remnant. A shorter distance and youngest age of HESS J1731-347 favors the latter scenario.

One may also consider the possibility that hadronic processes dominate the TeV emission. Given the hard electron spectrum inferred from the leptonic model, a hard proton spectrum is possible and compatible with the GeV upper limits. The hadronic model requires an even stronger magnetic field, however, which implies shorter energy loss time of X-ray emitting electrons and higher shock speed, which may pose a challenge given the large size at a distance of 3.2 kpc.

Finally, we note that the correlation among images of HESS 1731-347 made at different wavelengths does not appear to be as good as the other two sources, especially the absence of X-ray emission to the west of the remnant where both radio and TeV emission have been detected  \citep{hess2011}. \citet{Bam2012} even treated this region as a background when analyzing the {\it SUZAKU} X-ray observations. It is possible that regions of strong magnetic field with a small volume filling factor dominate the synchrotron emission while most of the TeV comes from larger regions with a relatively weaker magnetic field, which reduces the total energy content of the magnetic field.
Since the synchrotron emissivity is proportional to the magnetic field energy density, a relatively small amount of energetic electrons are needed to produce the X-ray spectrum.
In a simple two-zone model, there is a reservoir of energetic electrons with a weak magnetic field that produces the observed TeV emission. The X-ray is produced by energetic electrons entering a region of strong magnetic field. The remnant is then in a particular stage when the electron acceleration may have stopped and the total energy of energetic electron is decreasing with time. The decay timescale is given by the cooling time of electrons producing the observed X-ray via synchrotron process in the region of strong magnetic field divided by the fraction of energetic electrons in this strong field region. For a given synchrotron spectrum, it is proportional to $B^{-0.5}$. We estimate a decay timescale of a few thousand years. This may also explain the relatively lower total energy of energetic electrons in this source. Further X-ray observations are warranted.

 Recent detailed studies of RCW 86 are very revealing in this regard \citep{2012A&A...545A..28L}.  The total energy of electrons above 1 GeV is also much less than the total energy of the magnetic field in this SNR (Table 3).
Compared with RX J0852.0-4622 and RX J1713.7-3946, the magnetic energy of this source is very high. Hard X-ray and radio images clearly show that the synchrotron emission regions have a volume filling factor much less than 0.88 for the TeV emission region, implying a lower value for the total magnetic field energy. The mismatch between the radio and X-ray images implies spatial variation of the electron distribution. 
Therefore, in light of the spatial inhomogeneity of energetic electrons and/or magnetic field in this source, the energy density of energetic electrons and magnetic field may still be equal locally. Advanced modeling taking of these inhomogeneities and the electron energy loss into account is warranted.


\section{Conclusion}

In this paper, we carried out detailed analysis of four-year observations of the {\it Fermi}-LAT for HESS J1731-347 and found no excess flux correlated with the HESS source in the residual map above 2 GeV. Upper limits in the GeV band were then derived, which rules out simple hadronic models with a proton spectral slope greater than 1.8. These results further confirm that HESS J1731-347  is very similar to the TeV-bright shell-type
SNR RX J1713.7-3946, RCW 86, and the Vela Junior SNR RX J0852.0-4622 representing a unique population of SNRs with emission likely dominated by energetic electrons accelerated by shocks of the remnants \citep{Y2012}.

A detailed comparative study of model parameters in the leptonic scenario for these sources, however, shows that HESS J1731-347 might be distinct from RX J1713.7-3946 and RX J0852.0-4622, caused mainly by its high X-ray to $\gamma$-ray flux ratio, which leads to a very high magnetic field stress. {  Assuming a distance of 1 kpc}, SNR RX J1713.7-3946 and the Vela Junior SNR RX J0852.0-4622 shows energy equipartition between electrons above 1 GeV and the magnetic field in the emission region. Their difference may be attributed to a lower density in RX J0852.0-4622. To keep this energy equipartition in HESS J1731-347, the remnant needs to be within 500 pc from the Earth, implying a very high shock speed of $\sim 1\arcsec$ per year that may be tested with Chandra observations. Although such a short distance may bring HESS J1731-347 into a uniform model with the other two, where the mean magnetic field decreases monotonically with the remnant expansion, it is challenged by the high absorption column depth derived from the X-ray spectral measurement \citep{Bam2012, hess2011}.

Alternatively, the magnetic field in HESS J1731-347 may indeed have a much higher total energy than energetic electrons, the electron acceleration is then affected by the environment of the remnant significantly and a lower shock speed is expected. But the shock speed still needs to be a few thousand km/s in the one zone model given the short energy loss time of electrons producing the observed X-ray. Such a constraint does not exist for models with a complex source structure as suggested by differences in images at different wavelengths { and exemplified recently by detailed studies of RCW 86 \citep{2012A&A...545A..28L}}. Measurement of the shock speed with high resolution X-ray images and more comprehensive multiwavelength observations will be able to distinguish these scenarios, which will deepen our understanding of electron accelerations by shocks of SNRs dramatically.



\section*{Acknowledgements}
S.L. thanks Dr. Tian, W. W. for helpful discussions on the distance measurement.
This work is partially supported by the 973 grants 2013CB837000 and 2009CB824800, the Strategic Priority Research Program - The Emergence of Cosmological Structures of the Chinese Academy of Sciences, Grant No. XDB09000000, the NSFC grants 11105155, 11173064, 11233001, and 11233008, grant 20120091110048 from the Educational Ministry of China, and grants from the 985 Project of the NJU and the Advanced Discipline Construction Project of Jiangsu Province.

\bibliographystyle{aa}
\bibliography{ms-4}

\end{document}